\documentclass{ws-procs9x6}

\usepackage{epsfig}
\usepackage{rotate}

\newcommand{\mct}{{\mathcal T}}
\newcommand{\mcv}{{\mathcal V}}
\newcommand{\mi}{\mathrm i}
\newcommand{\foh}{\frac{1}{2}}
\newcommand{\fth}{\frac{3}{2}}
\newcommand{\ffh}{\frac{5}{2}}
\newcommand{\ra}{\rightarrow}

\begin{document}

\title{Vector meson production and nucleon resonance analysis in the
  Giessen coupled-channel model}

\author{\underline{G. PENNER}$^*$ and U. MOSEL}

\address{
Institut f\"ur Theoretische Physik, Universit\"at Giessen,
  D-35392 Giessen\\ 
$^ *$E-mail: gregor.penner@theo.physik.uni-giessen.de
}

\maketitle

\abstracts{
We present a nucleon resonance analysis by simultaneously considering
all pion- and photon-induced experimental data on the final states
$\gamma N$, $\pi N$, $2\pi N$, $\eta N$, $K\Lambda$, $K\Sigma$, and
$\omega N$ for energies from the nucleon mass up to $\sqrt s = 2$
GeV. In this analysis we find strong evidence for various higher lying 
resonances as, e.g., the $P_{13}(1900)$. The $\omega N$ production mechanism
is dominated by large $P_{11}(1710)$ and $P_{13}(1900)$ contributions.}

\section{Introduction}

The reliable extraction of nucleon resonance properties from experiments 
where the nucleon is excited via either hadronic or electromagnetic 
probes is one of the major issues of hadron physics. The goal is 
to be finally able to compare the extracted masses and partial-decay widths 
to predictions from lattice QCD and/or quark models (see references in Refs. \cite{penner2,penner3,phd}).

Basically all information about nucleon resonances identified so far
from experiment stems from analyses of pion-induced $\pi N$ and $2\pi N$ 
production and from pion photoproduction, see \cite{pdg} and 
references therein. However, it is well known 
that, for example, in the case of the $S_{11}(1535)$ the consideration
of the $\eta N$ final state is inevitable to extract its properties
reliably, and similar effects can be expected for higher-lying
resonances and different thresholds. 
On the other side, quark models predict a much richer resonance spectrum 
than has been found in $\pi N$ and $2\pi N$ production so far, giving
rise to speculations that many of these resonance states only couple
weakly to the $\pi N$ final state (``missing'' resonances). 
For a consistent and reliable identification of these resonances and 
their properties, the consideration of unitarity effects are
inevitable and as many final states as possible have to be taken into
account simultaneously. With this aim in mind we have developed in Refs. 
\cite{penner2,penner3,feusti,penner1} a unitary coupled-channel
effective Lagrangian model (\textit{Giessen  model}) that 
incorporates the final states $\gamma N$, $\pi N$, $2\pi N$, $\eta N$, 
$K \Lambda$, $K \Sigma$, and in particular $\omega N$. This model is
used for a simultaneous analysis of all available experimental data on
photon- and pion-induced reactions on the nucleon in the energy range
$m_N < \sqrt s < 2$ GeV. The premise is to use the
\textit{same Lagrangians} for the pion- and photon-induced reactions
allowing for a consistent analysis, thereby generating the background
dynamically from Born, $u$- and $t$-channel contributions without new
parameters.

Here, we only point out some aspects of our results; all the details
can be found in Refs. \cite{penner2,penner3,phd}.

\section{The Giessen Model}

In the Giessen model the Bethe-Salpeter equation for the scattering
amplitude $\mct$ is solved by the $K$-matrix approximation, where the
intermediate particles are set on their mass shell. By a
partial-wave decomposition in total spin $J$, parity $P$, and isospin 
$I$, which is performed in 
a generalized and uniform way for all final-state combinations (see
Refs. \cite{penner2,penner3,phd} for details), 
one arrives at an algebraic equation relating the decomposed
$\mct_{fi}$ and the potential $\mcv_{fi}$ (in a schematic
notation): $\mct^{IJ\pm}_{fi} = 
[ \mcv^{IJ\pm} / (1 - \mi \mcv^{IJ\pm}) ]_{fi}$. 
Hence unitarity is fulfilled as long as $\mcv$ is
Hermitian. Furthermore, in the effective Lagrangian approach
generating the potential $\mcv$, gauge invariance is fulfilled in
a straightforward way \cite{penner3,phd}.

\begin{figure}[t]
  \begin{center}
    \parbox{115mm}{
      \parbox{57mm}{\includegraphics[width=57mm]{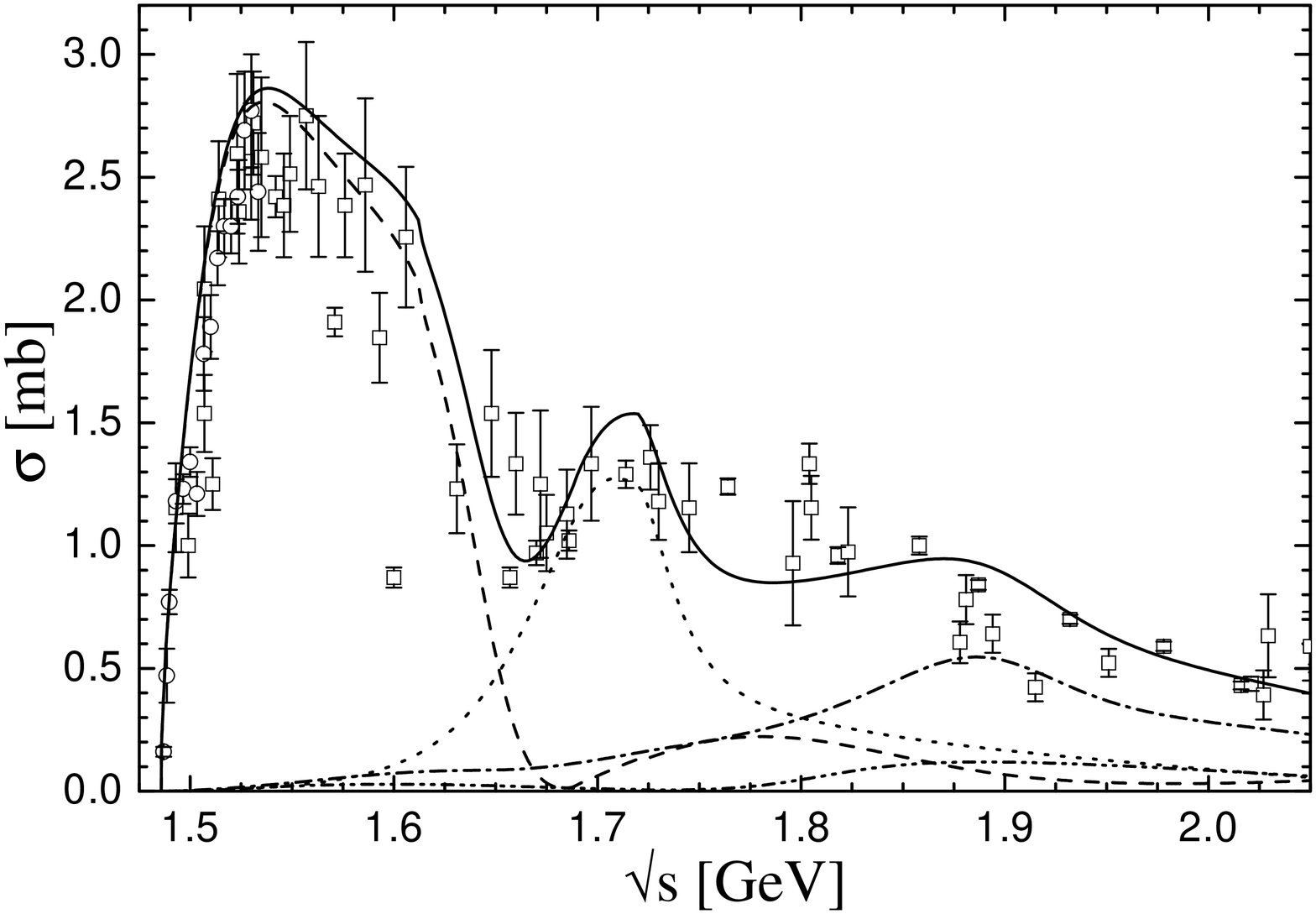}}
      \parbox{57mm}{\includegraphics[width=57mm]{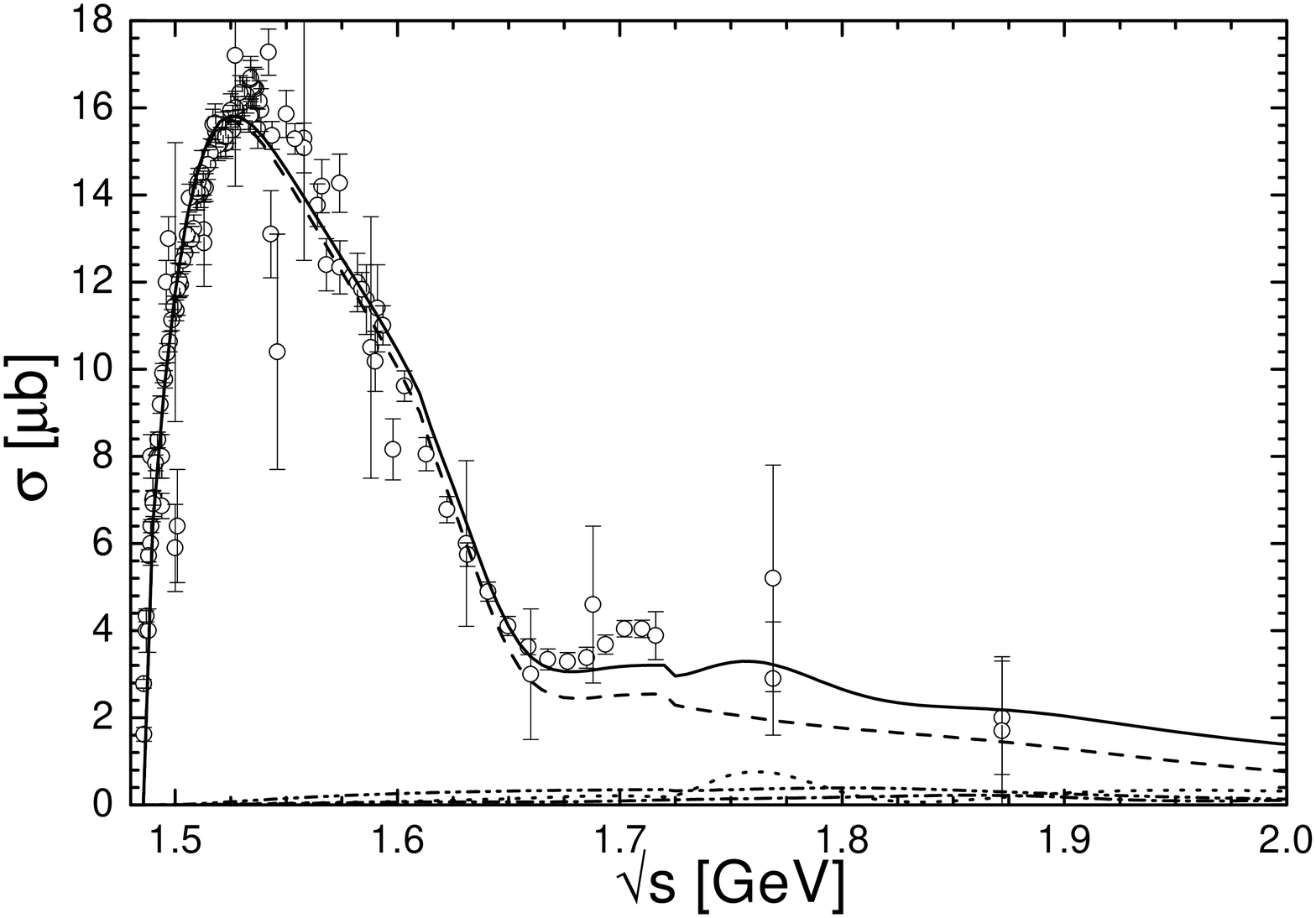}}
      }
    \parbox{115mm}{
      \parbox{57mm}{\includegraphics[width=57mm]{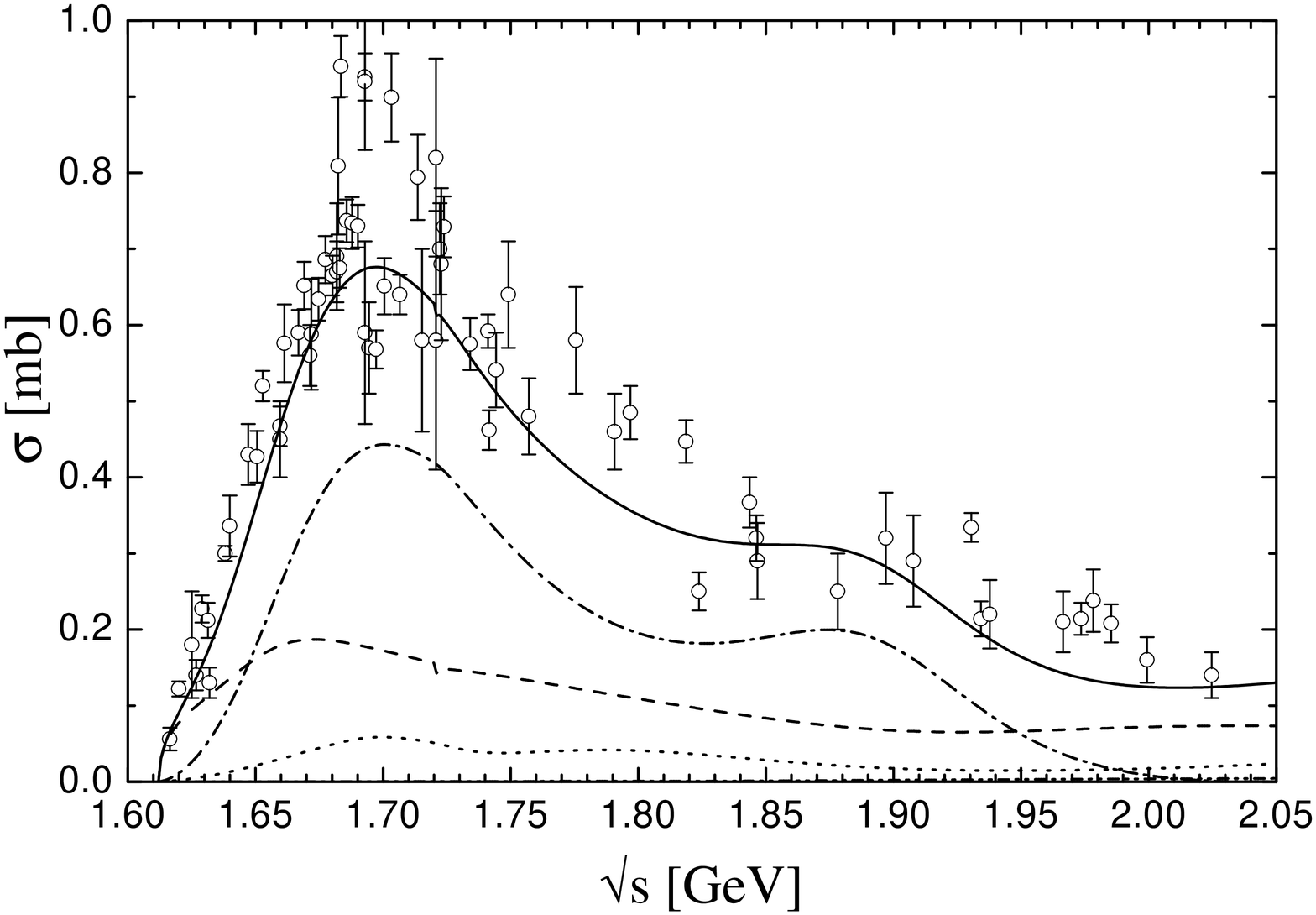}}
      \parbox{57mm}{\includegraphics[width=57mm]{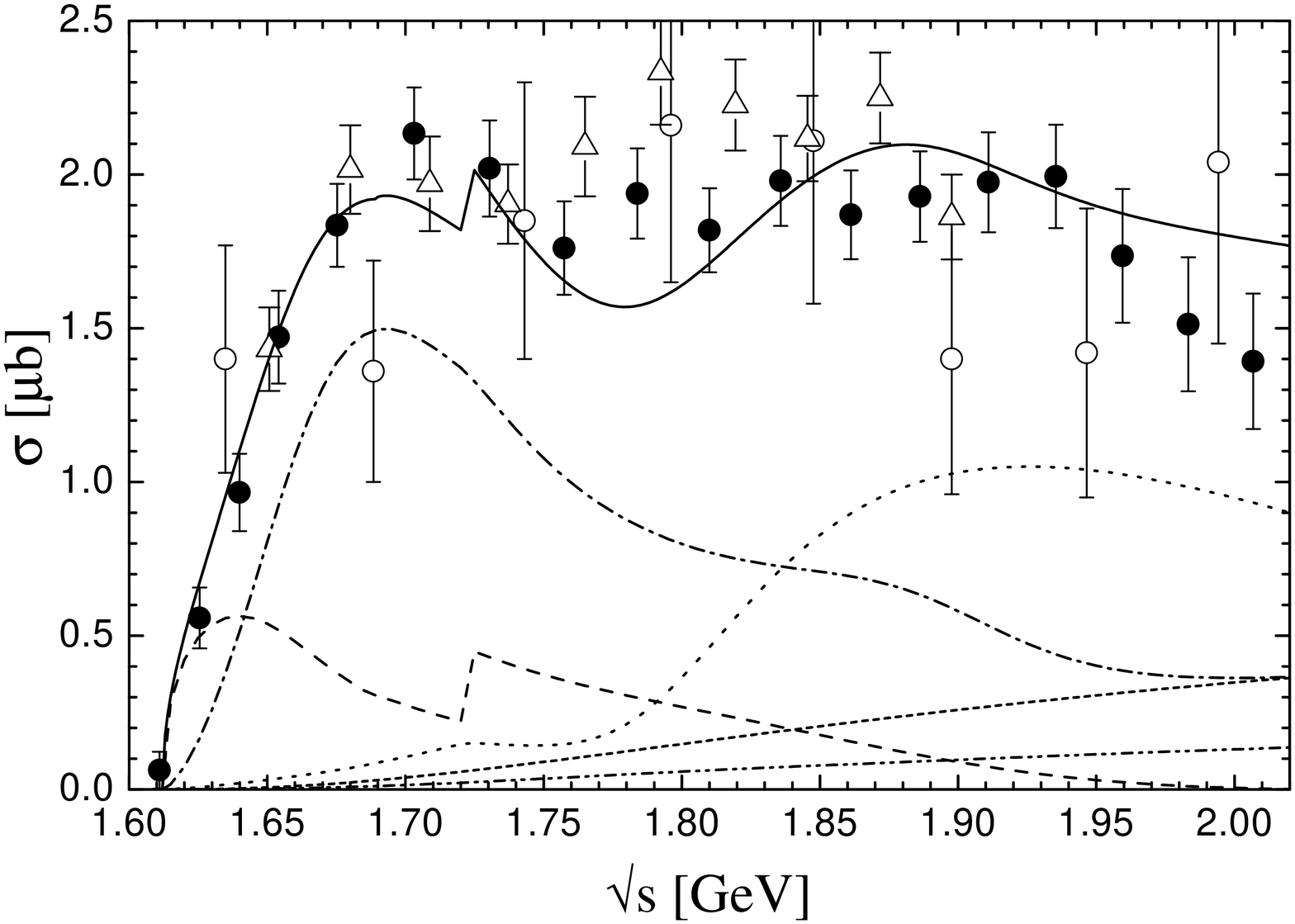}}
      }
    \parbox{115mm}{
      \parbox{57mm}{\includegraphics[width=57mm]{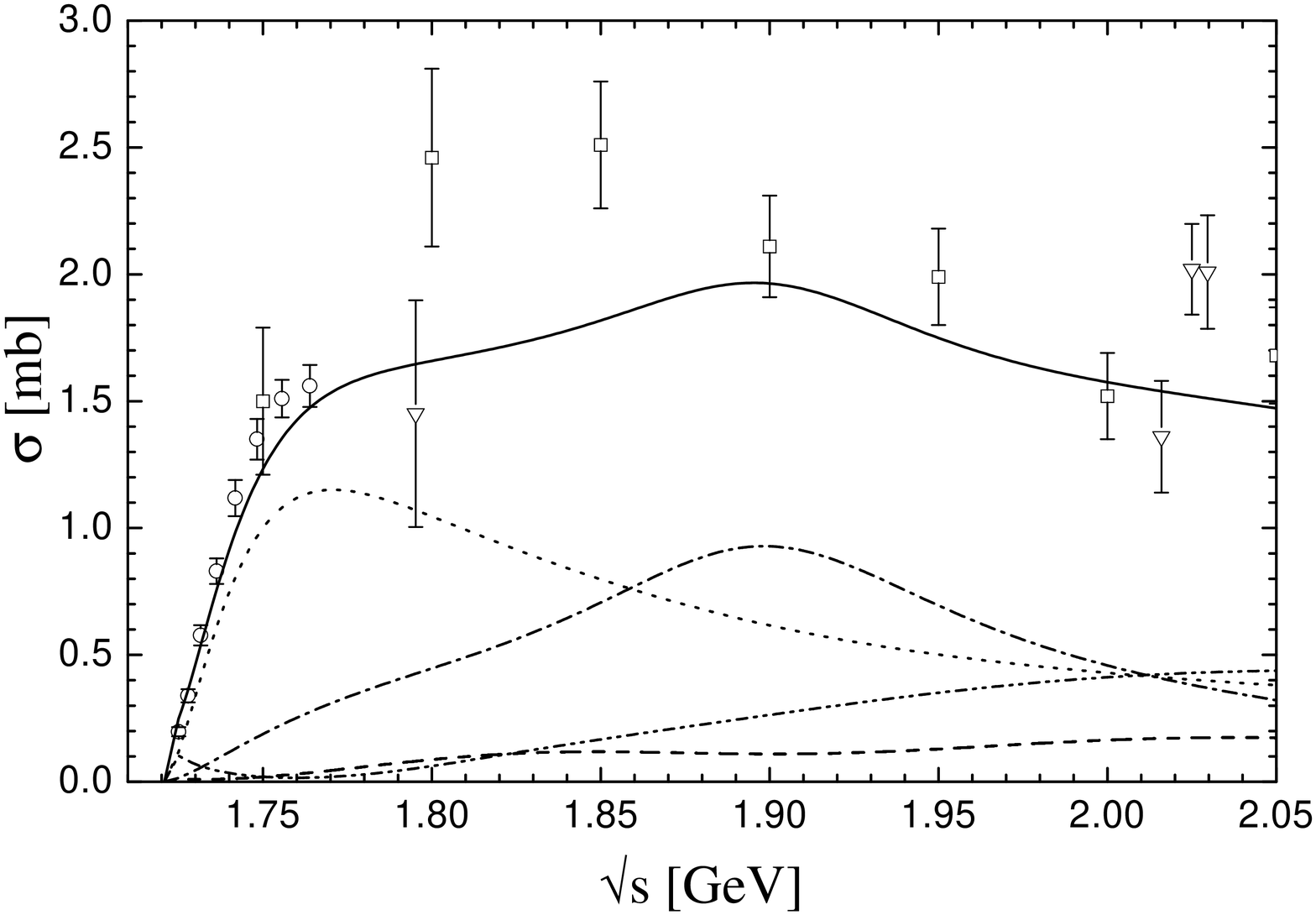}}
      \parbox{57mm}{\includegraphics[width=57mm]{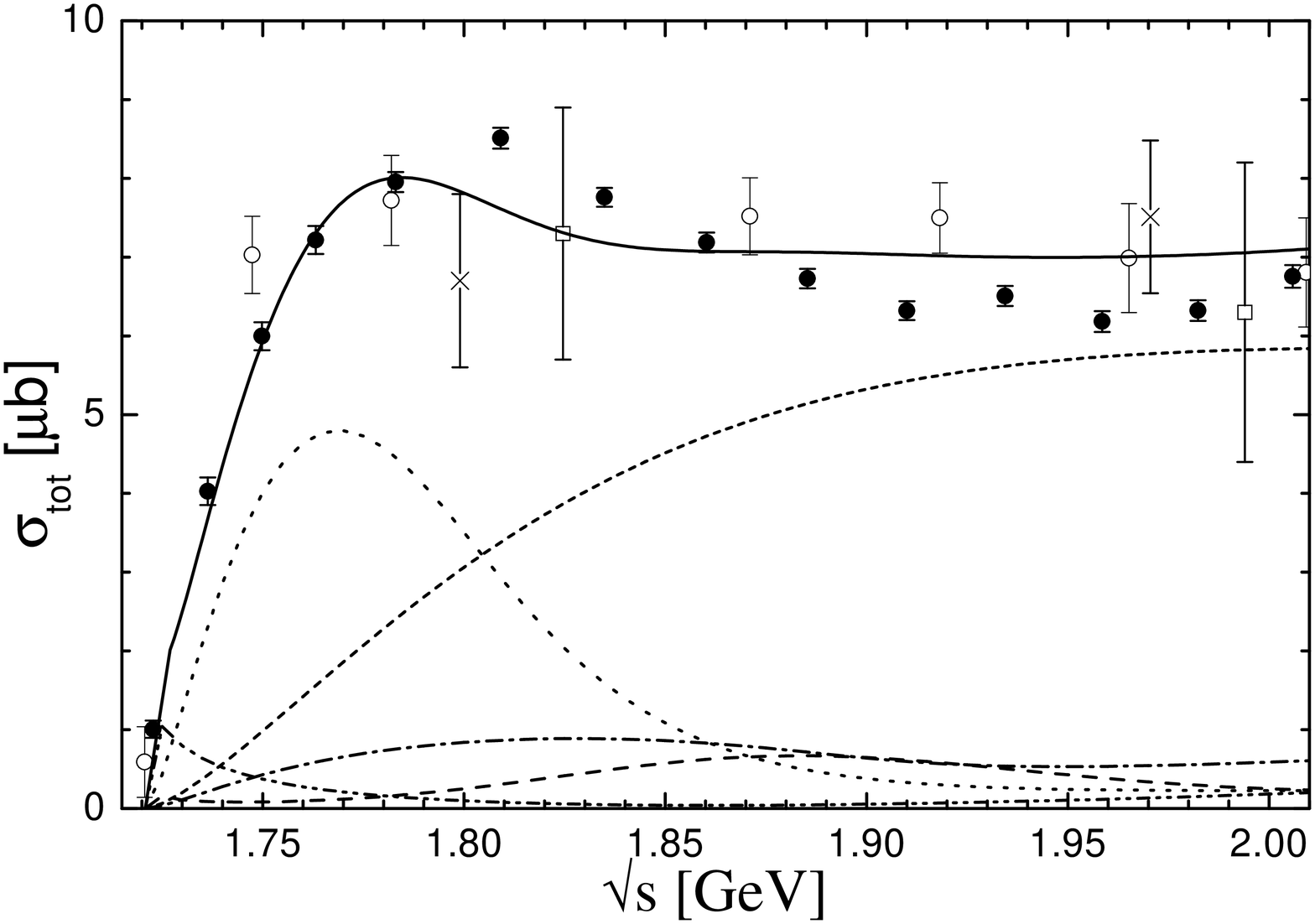}}
      }
    \caption{Examples of total cross section results. Upper left: 
      $\pi^- p \ra \eta n$, upper right: $\gamma p \ra \eta p$, 
      middle left: $\pi^- p \ra K^0 \Lambda$, 
      middle right: $\gamma p \ra K^+ \Lambda$, 
      lower left: $\pi^- p \ra \omega n$, 
      lower right: $\gamma p \ra \omega p$. 
      Partial-wave contributions: 
      $J^P=\foh^-(S_{11})$: dashed line; $\foh^+(P_{11})$: 
      dotted line; $\fth^+(P_{13})$: dash-dotted line;
      $\fth^-(D_{13})$: dash-double-dotted line (in brackets the $\pi
      N$ notation is given). The sum of all partial waves is given by
      the solid line. For
      $K\Lambda$ and $\omega N$ photoproduction, also the
      contribution of higher partial waves ($J\geq \ffh$) is 
      indicated by the short-dashed line. For the data references, see
      Refs. $^{1,2,3}$.
      \label{figtotals}}
  \end{center}
\end{figure}

\begin{figure}[t]
  \begin{center}
    \parbox{115mm}{\includegraphics[width=115mm]{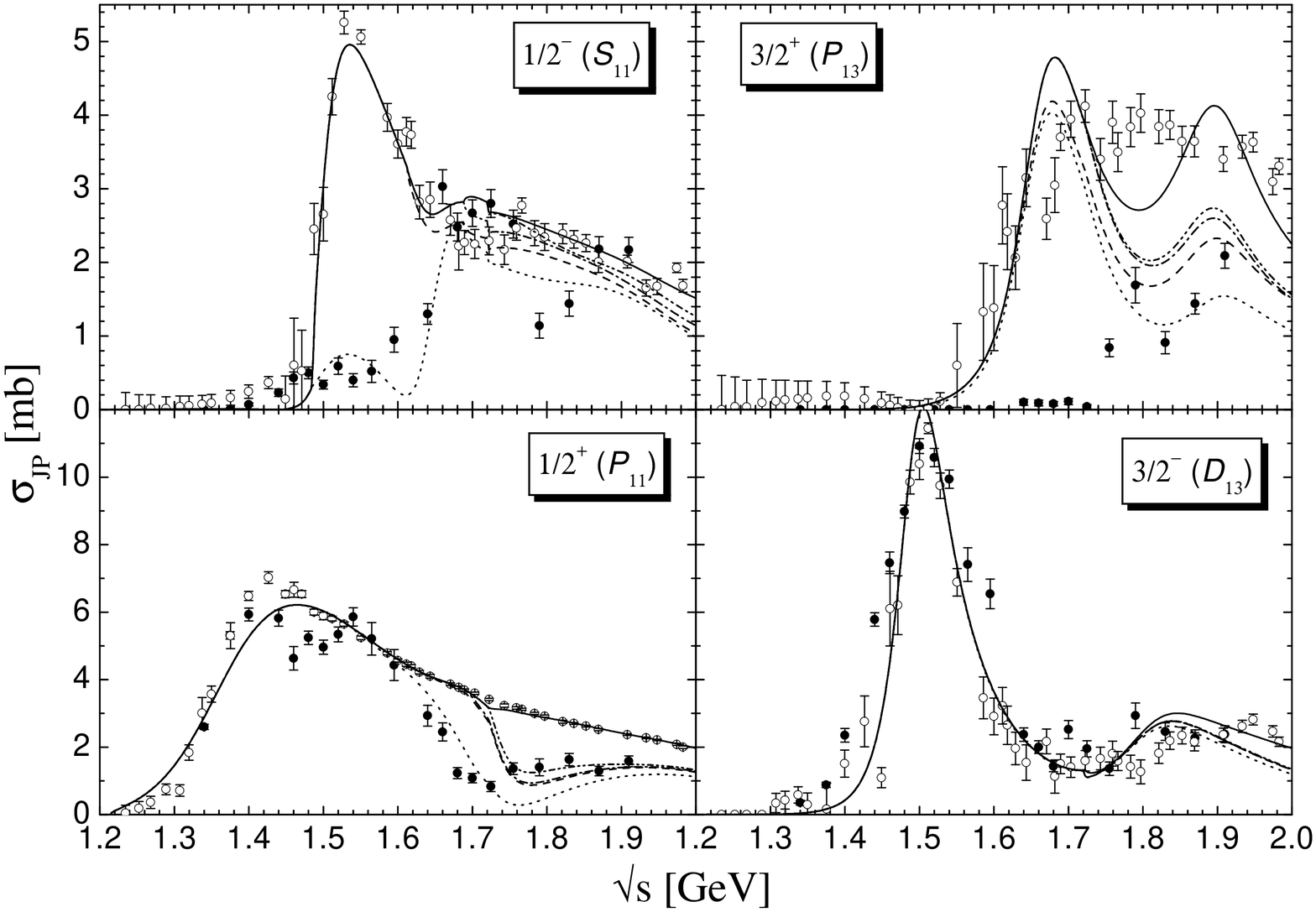}}
    \caption{Decomposition of the inelastic partial-wave cross
      sections of $\pi N \ra \pi N$ for isospin $\foh$. Partial-wave 
      cross section of
      $2\pi N$: dotted, $+\eta N$: dashed, $+K\Lambda$: dash-dotted,
      $+K\Sigma$: dash-double-dotted, total ($+\omega N$): solid
      line. The solid dots ($\bullet$) are $2\pi N$ partial-wave cross 
      sections, the open dots ($\circ$) are inelastic $\pi N \ra
      \pi N$ partial-wave cross sections. For the data references, see
      Refs. $^{1,2,3}$.
      \label{figinelast}}
  \end{center}
\end{figure}

The nucleon resonance properties are extracted by a $\chi^2$
minimization procedure using a total of about 6800 experimental data points,
comprising partial-wave analyses for $\gamma/\pi N \ra \pi N$,
$\pi N \ra 2\pi N$, and total and differential cross section
and polarization data for all other reactions (see
Refs. \cite{penner2,penner3,phd}); a good simultaneuos description of all
channels is achieved. As examples, we show the resulting total cross
sections of $\eta N$, $K\Lambda$, and $\omega N$ production in
Fig. \ref{figtotals}.
In these reactions and in particular in $\omega N$ production, strong
evidence for important $P$-wave contributions is found, mainly steming from
the $P_{11}(1710)$ and $P_{13}(1900)$ resonances, where the latter
one is especially 
important in the pion-induced reactions. In $K\Lambda$
photoproduction also a clear threshold effect is visible due to the
opening of the $\omega N$ final state. Note that in $\omega N$
photoproduction, we furthermore observe a large contribution of the
$\pi$ $t$-channel exchange. In Fig. \ref{figinelast}, 
the $\pi N$ inelastic partial-wave cross sections for $I=\foh$ are
shown, emphasizing the importance of the inclusion of final states
apart from $\pi N$ and $2\pi N$ in a unitary model. In the
$J^P=\foh^-(S_{11})$, $\foh^+(P_{11})$, and $\fth^+(P_{13})$ partial
waves the consideration of the $\eta N$ and $\omega N$ final 
states are particularly important.

\section{Outlook}

The next step towards a further improvement of the model is the inclusion of
spin-$\ffh$-resonance effects, see Ref. \cite{vitalyi}. Furthermore,
investigations concerning a detailed inclusion of all individual $2\pi 
N$ final states ($\rho N$, $\pi \Delta$, ...) are in progress.

\section*{Acknowledgments}
This work was supported by DFG and GSI Darmstadt.

\end{document}